\documentclass[12pt]{article}
\usepackage{graphics}
\usepackage{amsfonts}
\begin{document}

\title{A Method Based on a Nonlinear Gener
alized Heisenberg Algebra to Study the Molecular Vibrational Spectrum} 

\author{J. de Souza$^{(1)}$\\ 
 N. M. Oliveira-Neto$^{(1,2)}$\\
  C.I. Ribeiro-Silva$^{(1)}$\\
$^{(1)}$Centro Brasileiro de Pesquisas F\'{\i}sicas,\\
 22290-180, Rio de Janeiro - RJ \\
$^{(2)}$Universidade Federal de Vi\c cosa,\\
Departamento de F\'{\i}sica, 36570-000, MG, Brazil
}
\date{}

\maketitle
\begin{abstract}
\indent
We propose a method, based on a Generalized Heisenberg Algebra (GHA), to reproduce the anharmonic spectrum of diatomic molecules. The theoretical spectrum generated by GHA allows us to fit the experimental data and to obtain the dissociation energy for the carbon monoxide molecule. Our outcomes are more accurate than the standard models used to study molecular vibrations, namely the Morse and the $q$-oscillator models and comparable to the perturbed Morse model proposed by Huffaker \cite{hf}, for the first experimental levels. The dissociation energy obtained here is more accurate than all previous models.
\end{abstract}

\newpage

\section{Introduction}

Vibrational spectroscopy has a fundamental role in molecular physics and its applications extend to other fields like astronomy \cite{msk}, biology \cite{haken} and earth and environmental sciences \cite{fmw}. Vibrational molecular analysis provides important informations on the structure of the molecules. Recently, the development of the powerful experimental techniques which allow the study of highly excited vibrational states \cite{wang,ky}. In Born-Oppenheimer approximation, electronic, rotational and vibrational quantum states \cite{haken} can be considered separately and molecular vibrations can are approximately described by the harmonic oscillator \cite{cy}. However, this model fails to describe the highest molecular vibrational modes and consequently to provide the correct dissociation energies.  One of the first attempts to improve these approximations is due to Morse \cite{morse} who introduced a simple potential which allows to solve exactly the Schr\"odinger Equation (SE) and provides a reasonably approximation to the spectrum of diatomic molecules, including an upper bound which lacks in the harmonic oscillator potential. Later on, many other approximations based on the use of Morse-like, Kratzer-like  or modified versions of these or other potential functions have been introduced in order to obtain better fittings with experimental data \cite{partridge,dulick,selg,pliva} and  for evaluating dissociation energies \cite{ssh,dk}. 

Although {\it ab initio} methods have shown to be able to provide good approximations for the spectra of many molecules, some authors have pointed out some of their limitations. Angelova {\em et al.} \cite{angelova2}, noted that the Morse potential contains only quadratic corrections and one needs to use the empirical Dunham expansion to fit the highest vibrational energies.  Iachello and Levine \cite{if}, remarked that the solution of the SE is very difficult in the case of two- and three- dimensions and, at this point, algebraic methods could bring some advantages. Problems related to the dissociation energy estimation and also a algebraic energy method to calculate it, were described by Sun {\em et al.} \cite{sun}. Nevertheless, besides the facts above mentioned, there are still some molecules whose potentials deviate strongly from Morse-like potentials \cite{witek}.

The algebraic approach to the study of molecular vibrational and rotational spectra was pioneered by F. Iachello in \cite{iachello} (for a recent review see \cite{io}). The method has been proved to be useful for calculating accurately highly excited vibrational levels and for describing the whole spectrum of complex molecules, while maintaining its simplicity in cases where the use of {\it ab initio} methods are not feasible in practice \cite{wang,xhz,lf,zd}. 

Deformed Heisenberg Algebras (DHA or q-oscillators) have attracted considerable attention since they have been proposed \cite{macfarlane,biedenharn}, due mainly to their potential physical applications \cite{paper2} (see ref. \cite{bonatsos} for a review). They are non-trivial generalizations of the Heisenberg Algebras characterized by the deformation parameter $q$. It is interesting to notice that for some values of the $q$-deformation  parameter, namely for $0<q<1$, the energy spectrum associated with the Deformed Heisenberg Algebras presents an upper bound, as in fact occurs in systems like composite particles \cite{qoscilador}, as for instance, diatomic molecules \cite{angelova2,bonats,angelova1,bd1}. Within this formalism $q$-oscillators are used as a model of anharmonic oscillators, and the value of $q$ is chosen to reproduce experimental molecular vibrational spectra. When compared with other methods, the DHA approach requires only a few parameters, allows analytical expressions  and fits well the first experimental levels. However, as it will become clear later, it does not provide always the correct dissociation energy because it may fail to fit the highest vibrational levels. Recently, many authors have discussed the physical interpretation of this parameter \cite{spiridonov,bdk,bd,sbgd} but this analysis is out of the scope of this work.

Recently, a Generalized Heisenberg Algebra (GHA), where the commutation relations among the operators depend on a characteristic function of the generalized number operator has been proposed \cite{paperdeles1}. If the characteristic function is linear, with slope $q^2$, the algebra corresponds to the $q$-oscillator algebra. In the present work we introduce and implement a nonlinear GHA which is able to describe typical features of the vibrational molecular spectrum  of the carbon monoxide molecule ($\mathrm{CO}$). The spectrum generated by this nonlinear GHA (nl-GHA) allows us to fit the $20$ first vibrational transitions and to obtain the correct dissociation energy, providing better global results than the methods above mentioned. 

The paper is organized as follows. In section \ref{sessao2} we discuss the solution of the Schroedinger Equation (SE) with Morse potential . In section 3 we introduce the GHA method. In section 4 we apply our GHA model to the $\mathrm{CO}$ molecule. Finally, section 5 is devoted to discussions and conclusions.

\section{Morse Potential for Diatomic Molecules: Vibrational Levels}
\label{sessao2}
Let us now turn our attention to one of the most familiar quantum descriptions of diatomic molecules. For low energy levels, the molecular potential can be approximated by the harmonic potential, whereas for high energy this approximation breaks down. In fact, the higher vibrational transitions exhibit a certain degree of anharmonicity. Thus, to take into account  anharmonic terms and rotational effects, nonlinear terms must be introduced in the energy expression \cite{haken}:
\begin{equation} \label{eq:11}
E_{\nu,J}=\sum_{l,m}=Y_{l,m}\left( \nu+\frac{1}{2} \right)^l \left[ J(J+1) \right]^m, 
\end{equation}
where $\nu$ and $J$ are the vibrational and the rotational quantum numbers, respectively. The above expression was obtained by Dunhan, using the WKB method \cite{hf}.

\begin{figure*}
\vspace*{5cm}   
\includegraphics{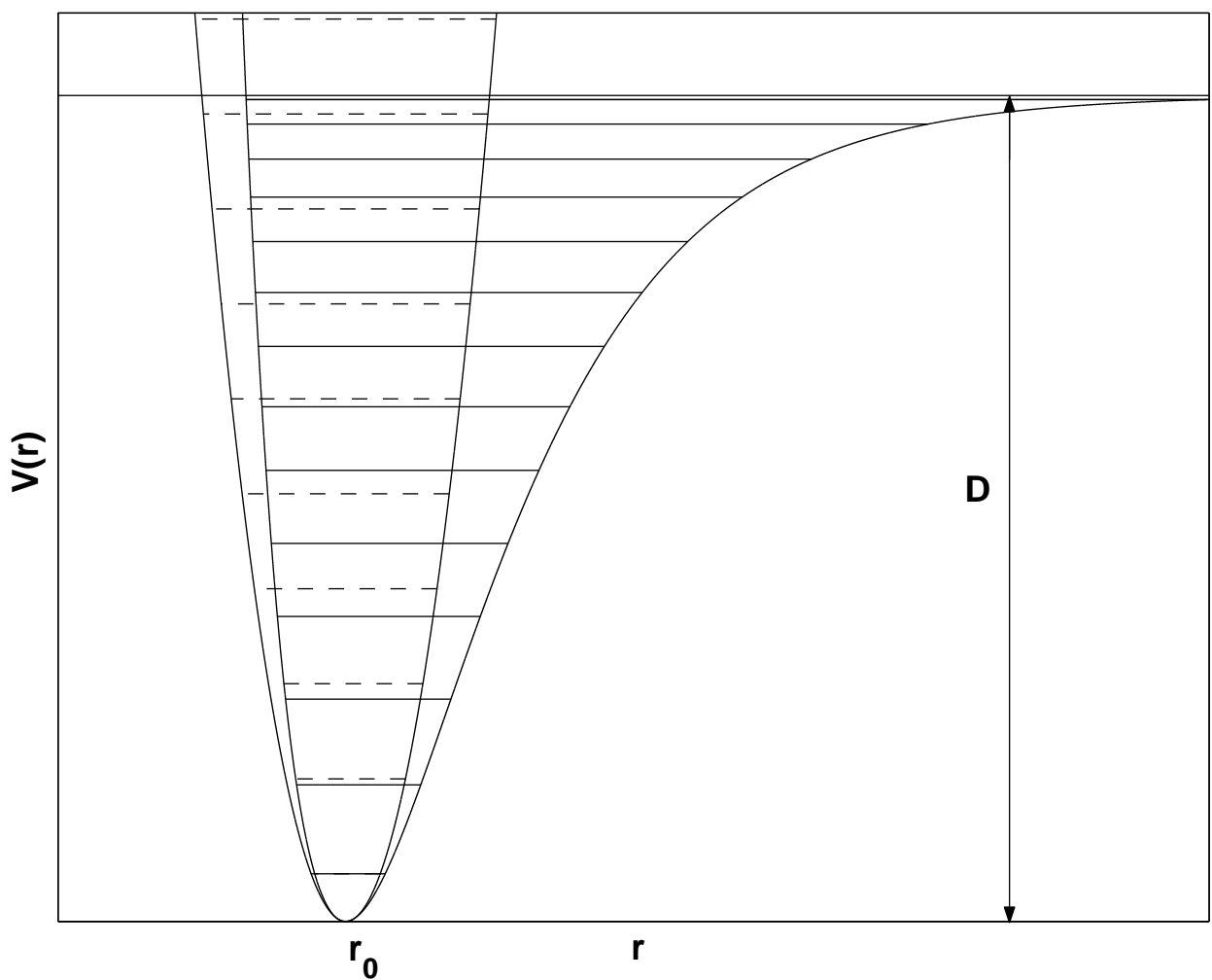}
\caption{Morse (full lines) and harmonic potential (dashed lines) energy levels. The horizontal axis is the relative nuclear position $r$. Note that the two potential have a minimum value in $r=r_0$ and that the Morse Potential has a dissociation energy $D$.}
\label{morsepot}
\end{figure*}

An analytical solution for the SE, which corresponds to the second order truncation of the equation (\ref{eq:11}) with $J_0=0$, was obtained by Morse \cite{morse} using the so-called Morse potential (see Figure \ref{morsepot}):
 \begin{equation} 
V(r)=De^{-2a(r-r_0)}-2De^{-a(r-r_0)} \label{eq:12},
\end{equation}
where $a$ is a characteristic constant of each molecule, $r$ is the nuclear distance, $r_0$ is the equilibrium position and $D$ is the dissociation energy. The solution of the SE for this potential gives the following expression for the vibrational energy 
\begin{equation} \label{eq:13}
\epsilon_n=-D+h\omega[(\nu +\frac{1}{2})-\chi_e(\nu +\frac{1}{2})^2],\, (\nu=0,1,...,\frac{k-1}{2}) ,
\end{equation}
where $\chi_e$ is a positive constant called the anharmonicity coefficient, $\omega$ is the fundamental 
frequency, $\nu$ is the vibrational quantum number and $k=\frac{1}{\chi_e}>1$. The  anharmonicity coefficient is related to the dissociation energy by, 
\begin{equation} 
\chi_e=\frac{h \omega}{4D}\label{eq:13a}.
\end{equation}
 
The energy levels obtained agree reasonably well with experimental data for many molecules \cite{haken}. The maximum  allowed value of $n$ in equation (\ref{eq:13}) is such that the SE solution remains finite. One can also show that this value is given by
\begin{equation} 
\nu_{max}=\frac{1}{2 \chi_e}-\frac{1}{2}.\label{eq:13b}
\end{equation}
Hence, there is maximum energy value $\epsilon_{\nu_{max}}$ beyond which the system is not bounded. 

In order to improve the Morse results, several perturbative methods, based on the Morse or other analytical or numerical potential, have been proposed. Huffaker \cite{hf}, used a perturbed Morse model to calculate the first terms of the Dunhan expansion (\ref{eq:11}) for the $\mathrm{CO}$ molecule. He used a potential given by following expression, 
\begin{eqnarray} 
V(r)&=& V_e[ (1-e^{-aq})^2+b_4(1-e^{-aq})^4 \\
& & + b_5(1-e^{-aq})^5 + \dots ) ] \, .\nonumber \label{eq:11d}
\end{eqnarray}
Using $28$ spectral lines of the $\mathrm{CO}$ molecule \cite{mwaz}, he calculated $8$ parameters of the potential curve given by equation (\ref{eq:11d}). This model is good when compared with the model obtained via RKR method by Mantz {\em et al.} \cite{mwaz}. If one ignores the rotational degree of freedom ($l=0$),  the model has $7$ parameters.

\section{The Generalized Heisenberg Algebra}
\label{sessao3}
In \cite{paperdeles1} Curado e Rego-Monteiro proposed an algebra called GHA which is generated by three operators $J_0$, $A$ and $A^\dagger$, satisfying the following relations:
\begin{eqnarray}
J_0 \,  A^\dagger &=& A^\dagger \,  f(J_0), \label{eq:1}\\
A  \, J_0 &=& f(J_0) \,  A, \label{eq:2}\\
\left[ A , A^\dagger \right] &=& f(J_0)-J_0,  \label{eq:3}
\end{eqnarray} 
where $^\dagger$ is the Hermitian conjugate, $(A^\dagger)^\dagger =A$, $ J_{0}^\dagger=J_0$ and $f(J_0)$ is an analytical function of $J_0$, called the characteristic function. Assuming the existence of a vacuum  state represented by $|0\rangle$ ($A|0\rangle\equiv 0$), it can be shown  that 
\begin{eqnarray}
J_0 \, |m-1\rangle &=& f^{m-1}(\epsilon_0) \, |m-1\rangle, m=1,2... \label{eq:4}, \\
A^\dagger \, |{m-1}\rangle &=& M_{m-1} \, |m\rangle, \label{eq:5} \\
A \, |{m}\rangle &=& M_{m-1} \, |{m-1}\rangle, \label{eq:6}
\end{eqnarray}
where $M^2_{m-1} = f^m(\epsilon_0) - \epsilon_0$, $\epsilon_0$ is the lowest $J_0$ eigenvalue and $f^m(\epsilon_0)$ is the $m$-th iteration of the function $f$. This algebra describes a class of Heisenberg-like algebras of quantum systems, having energy eigenvalues given by
\begin{equation}
\epsilon_n = f(\epsilon_{n-1}) \label{eq:7},
\end{equation}
where $\epsilon_n$ and $\epsilon_{n-1}$ are two successive eigenvalues and are related with the energy for the system \cite{naza}. Therefore, the $J_0$ eigenvalues can be obtained iteratively and they can be upper bounded or not, depending on the chracteristic function, the values of the function parameters and the initial  value $\epsilon_0$. For each kind of function, the values of the parameters and the initial values  determine the existence or not of the fixed points, $\epsilon^\star = f(\epsilon^\star)$ \cite{paperdeles2}, and their  stability. As a consequence, different spectra are obtained for different functions and the eigenvalues behavior can be analyzed using dynamical systems techniques. For the linear case, $f(J_0) = q \, J_0 + s$, the equations (\ref{eq:1}-\ref{eq:3}) become
\begin{eqnarray}
\left[ J_0 , A^\dagger \right]_{q} &=& s \, A^\dagger,  \label{eq:8} \\
\left[ J_0 , A \right]_{q^{-1}} &=& - \frac{s}{q} \, A,\\
\label{eq:9}
\left[ A , A^\dagger \right] &=& (q-1) \, J_0 + s,  \label{eq:10}
\end{eqnarray}
where $\left[a,b \right]_q\equiv ab-q\,ba$ is the $q$-commutation of the operators $a$ and $b$. Equations (\ref{eq:8}-\ref{eq:10}) describe the one-parameter GHA studied in \cite{paperdeles1} and can be mapped into the $q$-oscillator algebra. If $q=1$, the standard Heisenberg algebra is recovered.

\begin{figure*}
\vspace*{5cm}   
\includegraphics{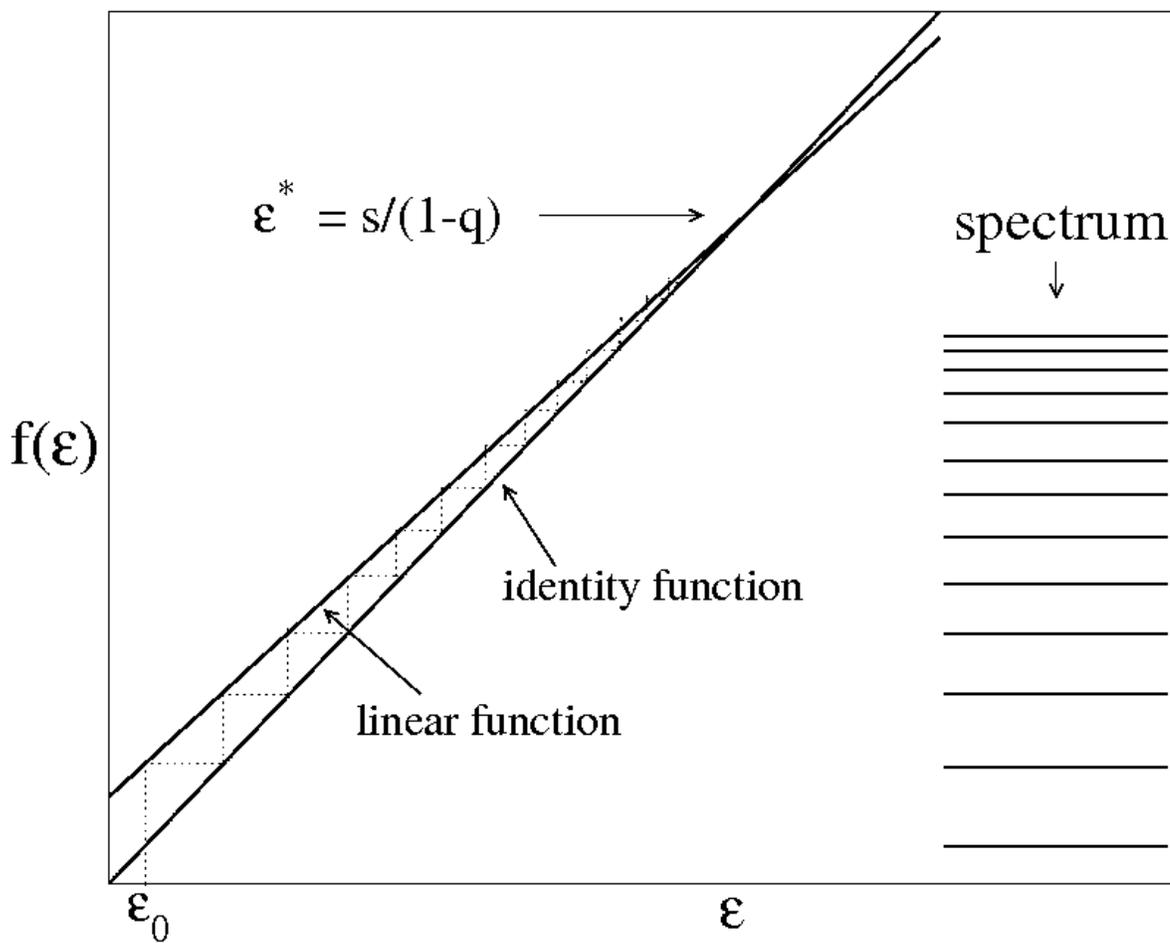}
\caption{Graphical analysis of a linear characteristic function, 
$f(\epsilon)$ for $0<q<1$ with $\epsilon_0<\frac{s}{1-q}$.
 Note that, as the function is iterated, the energy value approaches the stable point $\epsilon^*=\frac{s}{1-q}$ (the upper bound).  The dissociation energy of the system is proportional to $(\epsilon^*-\epsilon_0)$.}
\label{ga}
\end{figure*}

A graphical analysis of the functions $f(\epsilon)=q\epsilon+s$ and $y=\epsilon$ is shown in Figure \ref{ga}. The intersection between the two lines is identified as the fixed point of the recurrence equation $\epsilon_n = q\epsilon_{n-1}+s $. The most interesting cases are obtained for $0< q < 1$ and $\epsilon_0 <\frac{s}{1-q}$, since under these conditions the energy spectrum  has an upper bound $\epsilon^*=\frac{s}{1-q}$, as observed in the spectra of bounded systems, as for instance diatomic molecules \cite{wang,angelova2,bonatsos}.

\section{GHA and Vibrational Molecular Spectrum}
\label{sessao4}
\subsection{Linear case - ($q$-oscillator)}

We will now study the spectrum generated by the linear GHA mentioned above, which correspond to the $q$-oscillator algebra. To study the vibrational spectrum of diatomic molecules via GHA formalism, we use a result developed in \cite{paperdeles2,paperdeles3}. Let us start with the general Hamiltonian 
\begin{equation}\label{eq:13c}
H= \hbar\omega (c_1 AA^\dagger + c_2 A^\dagger A +c_3),  
\end{equation}
 where  $A$ and $A^\dagger$ obey the relations 
(\ref{eq:1}-\ref{eq:3}) and $c_1$, $c_2$ and $c_3$ are real numbers. Choosing $c_1=c_2=1$ and $c_3=0$, and using the equations (\ref{eq:4}-\ref{eq:6}) we get \cite{paperdeles2} 
\begin{equation}
H=\hbar\omega\left(f(J_0)+J_0-2
\epsilon_0\right). 
\end{equation}
Therefore, for the linear case $f(J_0)=qJ_0+s$ ($q<1$) we have:
\begin{equation}\label{eq:13d}
H= \hbar\omega\left((q+1)J_0+s-2
\epsilon_0\right). 
\end{equation}
Applying the Hamiltonian (\ref{eq:13d}) on the eigenstate $|{\nu}\rangle$ of $J_0$ (with $\hbar=1$) we obtain $H|{\nu}\rangle=E_{\nu}|{\nu}\rangle$,  where the energy eigenvalues are given by
\begin{equation}
E_\nu=\omega[(q+1)f^{\nu}(\epsilon_0)+s-2\epsilon_0], 
\end{equation}
and $f^{\nu}(\epsilon_0)=q^{\nu}\epsilon_0+s\frac{q^{\nu}-1}
{q-1}$. After some algebra we obtain 
\begin{equation}\label{eq:14-0}
E_{\nu}=\omega\left(M_q-L_q q^{{\nu}+1/2}\right),
\end{equation}
with $L_q=\frac{1+q}{q^{1/2}}\left(\epsilon_0-\frac{s}{1-q}\right)=\frac{1+q}{q^{1/2}}(\epsilon_0 -\epsilon^*)$ and $M_q=2(\frac{s}{1-q}-\epsilon_0)=2(\epsilon^* -\epsilon_0)$. 

The maximum energy ($\nu\rightarrow\infty$) for this system is given by $ E_{\infty}=\omega M_q=2\omega(\epsilon^* -\epsilon_0)$, and the dissociation energy is then 
\begin{equation}\label{eq:14}
D_q \equiv E_{\infty}-E_0 =\omega (1+q)(\epsilon^* -\epsilon_0).
\end{equation}
The $q-$parameter can be interpreted as being related to the anharmonicity constant, $\chi_e =-\frac{1}{2} \tau $, where  $ q=e^{\tau}$ \cite{bd}. 

We will now apply the previous results to the carbon monoxide molecule spectrum and compare it with both experimental data and the spectra obtained by using the Morse, the perturbed Morse and the $q$-oscillator models. We used the data provided by the HITRAN data\-base \cite{hitran} and chose the lines in which the molecules are in the electronic fundamental state. Also, we selected the lines whose rotational quantum numbers are zero. Thus, there remained the spectral lines which represent the first $21$ pure vibrational transitions of the $\mathrm{CO}$ molecule.

\begin{figure*}
\vspace*{5cm}   
\includegraphics{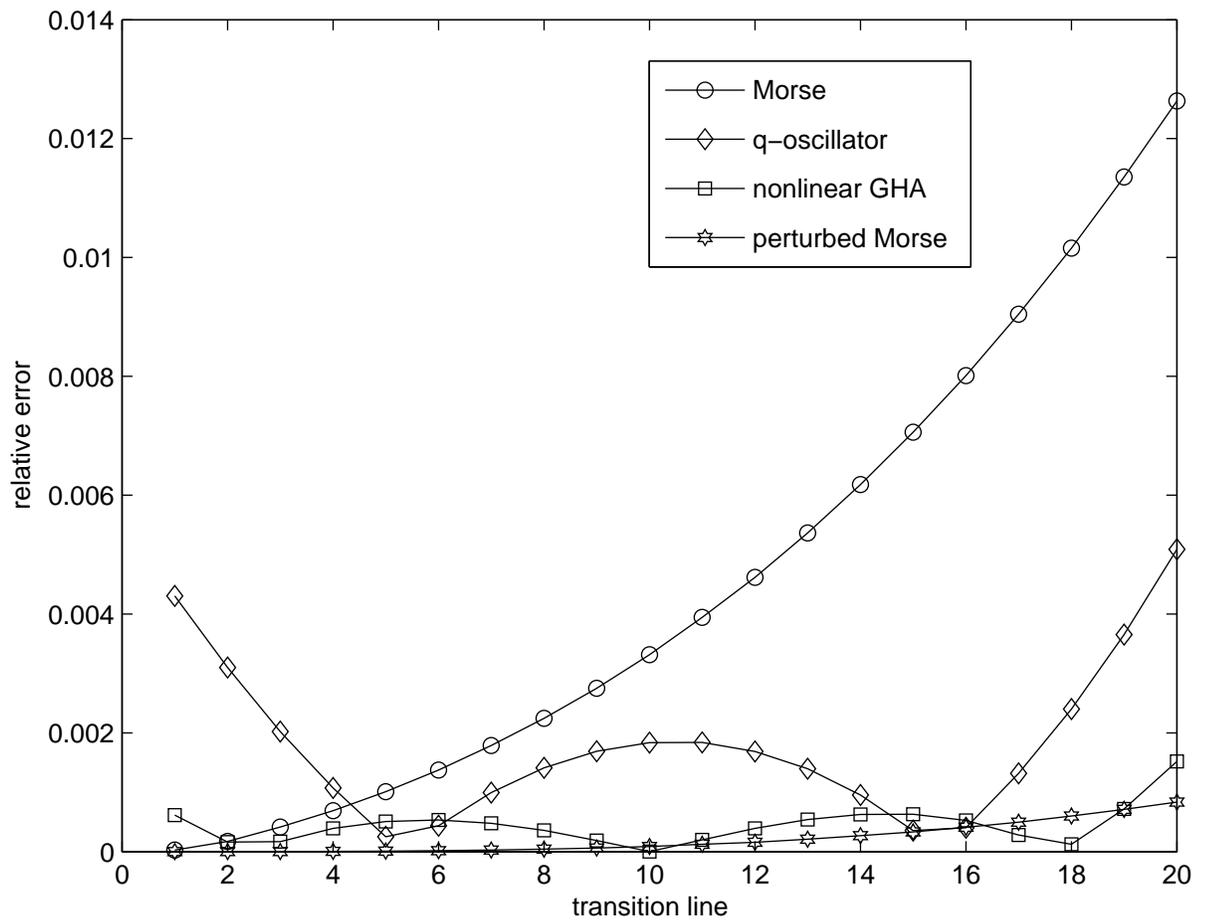}
\caption{Comparison between the relative errors of the Morse ($\circ $),   $q$-oscillator ($\diamond$), nonlinear GHA ($\Box$) models and perturbed Morse ($\star$). We can see that the nonlinear GHA errors are smaller (in almost all data points) than the Morse and $q$-oscillator models.}
\label{error}
\end{figure*}

In order to fit the parameters ($q$ and $\epsilon_0$) to the experimental data, note that the logarithm of the difference between two successive levels of the equation (\ref{eq:14-0}), as a function of $\nu$, is a straight line with slope $\ln q$. Taking the logarithm of the difference between two successive data points, we obtain the value of $q$ by evaluating the slope of the curve after a linear regression. The fitted parameters are shown in table (\ref{t1}). The linear GHA spectrum agrees well with the experimental data (see table (\ref{t2}) and figure (\ref{error})) mainly in the case of the $20$ first vibrational transitions. However, the $D$ value calculated by equation (\ref{eq:14}) deviate strongly from the experimental value as can be seen in table (\ref{t0}). This means that, although the $q$-oscillator relative errors ($\Delta E=\left|\frac{E_{theor.}-E_{exp.}}{E_{exp.}} \right|$) are smaller than the relatives error obtained by Morse model, for most of the 20 first vibrational transition (figure (\ref{error})), the approach based on the $q$-oscillator is not able to fit higher frequency vibrational levels. 

\subsection{Nonlinear case}
As we have seen above, the linear GHA ({\it i.e.} the q-oscillator), is a good model only for the first levels. In order to fit the energy levels and to obtain the correct dissociation energy we were led to use a nonlinear functional $f(x)$ in the GHA approach (\ref{eq:1}-\ref{eq:3}). We found that, by using the functional
\begin{equation}\label{eq:14a}
f(x)=p~x^4+q~x+1,
\end{equation}
it is possible to obtain a good fit with the experimental data and the correct dissociation energy. Due its nonlinearity, the spectrum generated by this nonlinear GHA does not have an analytical closed expression like equation (\ref{eq:14-0}). In this case the spectrum can only be calculated numerically. 

For the sake of simplicity, we start from the Hamiltonian 
\begin{equation}\label{eq:14a1}
H=\hbar w (A^\dagger A+\epsilon_0)= \hbar w J_0,
\end{equation}
which is derived from Hamiltonian (\ref{eq:13c}) with $c_1=0$, $c_2=1$ and $c_3=\epsilon_0$. Replacing $f(J_0)$ given by equation (\ref{eq:14a}) in relations (\ref{eq:1}-\ref{eq:3}) and applying $H$ on the eigenstates of $J_0$ (with $\hbar=1$)  we obtain 
\begin{equation}\label{eq:14a2}
E_\nu=w f^\nu(\epsilon_0),
\end{equation}
where  $f^{\nu}=\epsilon_{\nu}$ is given by:
\begin{equation}\label{eq:16}
\epsilon_{n+1}=p\epsilon_n^4+q\epsilon_n+1.
\end{equation}
We use the values generated by equation (\ref{eq:14a2}) with  $f^\nu(\epsilon_0)$ given by equation (\ref{eq:16}) in order to fit the $q$, $p$ and $\epsilon_0$ parameters with the experimental data. 
The dissociation energy is given by 
\begin{equation}\label{eq:15}
D_{GHA}=E_\infty-E_0= w(\epsilon^*-\epsilon_0),
\end{equation}
where $\epsilon^*$ is the stable fixed point of the recurrence equation (\ref{eq:16}).

The fitted parameters are shown in table (\ref{t1}). The dissociation energy are shown in table (\ref{t0}). In table (\ref{t2}) we compare the energy levels obtained with the different models referred in this work with experimental data. The relative errors are shown in the figure (\ref{error}). We can see that the nonlinear GHA with the functional (\ref{eq:14a}) provides better fittings with experimental data than both q-oscillator and Morse models and it is comparable to the Huffaker model \cite{hf}. Furthermore, nonlinear GHA provides a more accurate dissociation energy when compared with that obtained in all other methods. This results show that our model is a good method of obtaining the higher anharmonic energy levels of the $\mathrm{CO}$ molecule. 

\begin{figure*}
\vspace*{5cm}   
\includegraphics{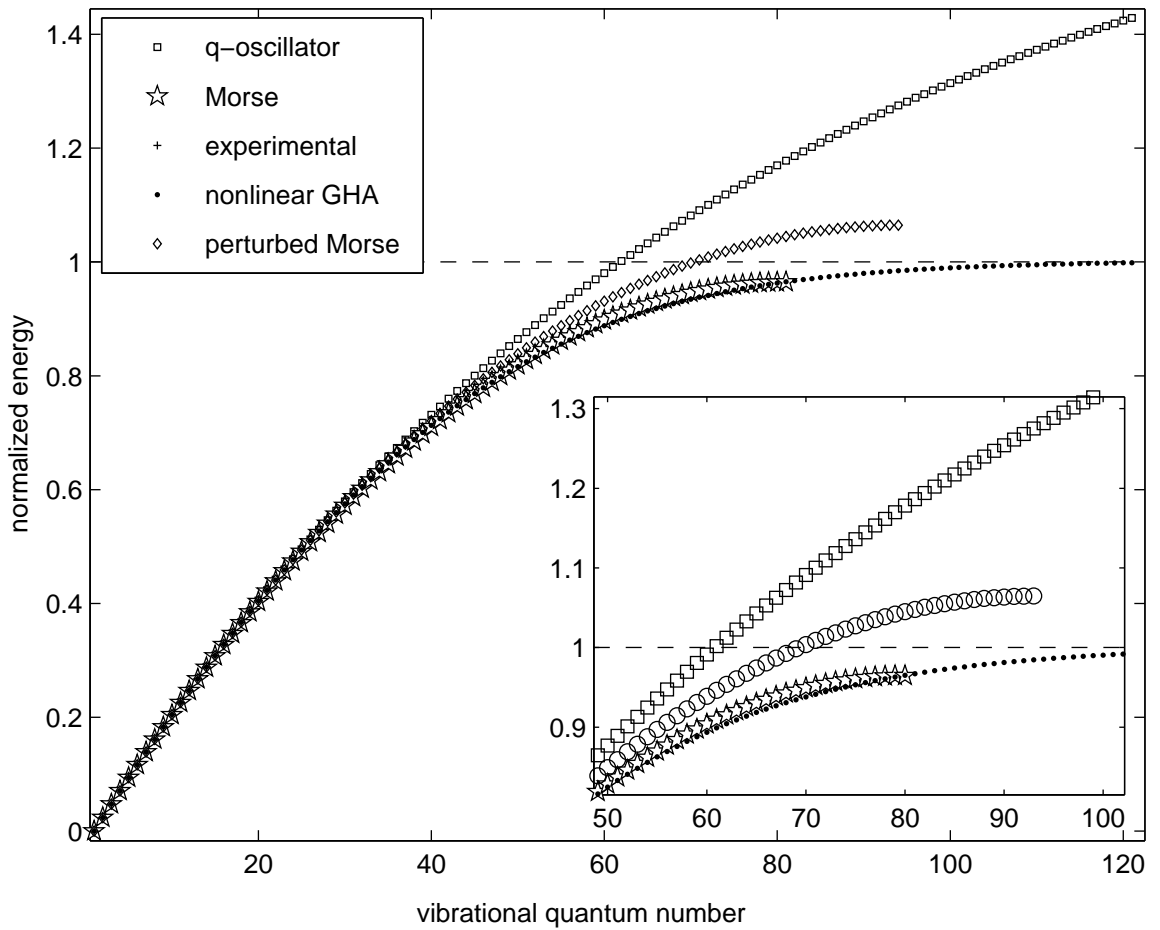}
\caption{Normalized and shifted energy spectrum generated by the Morse ($\star$), $q$-oscillator ($\Box$) and nonlinear GHA ($\cdot$) models, perturbe Morse ($\diamond$) and experimental data ($+$). The dashed line represents the experimental dissociation energy value. We can see that the dissociation energy of the nonlinear GHA lies on the experimental dissociation energy (dashed curve) while the $q$-oscillator and Morse spectrum are respectively  above and below dissociation energy line. Inset figure shows in detail the Morse and the nonlinear GHA energy curves close to dissociation energy line. Note that the Morse curve flats before reaching the experimental dissociation energy while the nonlinear GHA energy curve approach it as $\nu$ increase.}
\label{energy}
\end{figure*}

As the $p$ parameter is very small (table (\ref{t1})), if $\epsilon_0$ is also small, the first iterations of the equation (\ref{eq:16}) are dominated by the linear term. The nonlinear term becomes relevant as the number of iterations increase. The difference between two successive energy levels decreases up to zero as the function (\ref{eq:16}) is iterated. Because the $p$ parameter is negative, the difference between two successive energy levels tends to zero. Thus, the nonlinear term flats the energy curve faster than in the pure linear and the perturbed Morse cases but slightly slower than the Morse Model, as we can see in the figure (\ref{energy}).

We would like to stress that besides these accurate fittings, the nonlinear GHA gives us an extremely simple way to estimate higher transitions for the  $\mathrm{CO}$ molecule, just by iterating equation (\ref{eq:16}) up to the required level.

\section{Conclusion and Perspectives }

 In this work we have proposed a method based on GHA to reproduce the vibrational molecular spectrum of diatomic molecules. For the $\mathrm{CO}$ molecule, we have shown that the previous method based on the $q$-oscillator algebra (linear GHA) reproduce the vibrational molecular spectrum for the $20$ first vibrational transitions, but it does not provide the correct dissociation energy, {\it i.e.}, this model fails when describing higher vibrational levels.  Using a nonlinear functional ($4-th$ order) in the GHA we were able to fit the experimental energy levels and also to calculate the dissociation energy for the $\mathrm{CO}$ molecule with good accuracy. For this molecule the nonlinear GHA has provided better global results than the usual Morse, perturbed Morse models and $q$-oscillator model. This work shows that GHA can be used as a phenomenological simple tool to study composite particles. 

GHA produces different spectra for different characteristic functions. The GHA spectrum could be related to deformations in harmonic potential. Indeed, this algebra could reproduce a variety of spectra once we are able to find the appropriate characteristic function. Consequently, this algebra could be used to study other molecules, simply choosing the appropriate characteristic function. In an even more general fashion, de Souza {\em et al.} \cite{desouza} have constructed a  more general structure that depends on two functionals. This new algebraic structure can reproduce a even greater variety of spectra allowing us to work with two quantum numbers.

The results obtained in this work motivate us to further explore the relation between the GHA and molecular systems. This investigations may include: 1) the study of the relation between the functional parameters and the molecular parameters; 2) the investigation of the potential curve underling the GHA and 3) the account of electronic and rotational states.

\section*{Acknowledgments}
We thank E.M.F. Curado and M.A. Rego-Monteiro for fruitful discussions and remarks and Francisco Tamarit, S.M.D. Queir\'os, N. Lemke, E.R.P. Delfin , F.D. Nobre and F. Baldovin for manuscript revision. We would also like acknowledge the  partial support of CNPq and PRONEX (Brazilian agencies).

\begin{table}
\large
\centering
\begin{tabular}{|c|c|c|}\hline
model         & parameters   & values\\\hline
Morse         & $\chi_e$     &$0.0062$\\\hline
$q$-oscillator& $q$          &$0.98646$\\\cline{2-3}
 (linear GHA) & $\epsilon_0$ &$36.98$\\\hline
nonlinear GHA& $\epsilon_0$ &$0.9235$\\\cline{2-3}
              & $q$          &$0.9872$\\\cline{2-3}
              & $p$          &$-1.43\times 10^{-7}$\\\hline
perturbed Morse& $\sigma$     &$77.21317$\\\cline{2-3}
(Huffaker)     & $\tau$         &$83769.28cm^{-1}$\\\cline{2-3}
              & $b_4$          &$0.036067$\\\cline{2-3}
              & $b_5$          &$0.017505$\\\cline{2-3}
              & $b_6$          &$0.014945$\\\cline{2-3}
              & $b_7$          &$0.010770$\\\cline{2-3}
              & $b_8$          &$0.008142$\\\hline
\end{tabular}
\caption{Values of the parameters used in each model.}
\label{t1}
\end{table}

\begin{table}
\large
\centering
\begin{tabular}{|c|c|}\hline
model & Dissociation Energy ($cm^{-1}$)\\\hline
Experimental  &$89591.35$\\\hline
Morse         &$86426.44$\\\hline
$q$-oscillator&$158970.48$\\\hline
Perturbed Morse\footnote{ From reference \cite{hf}} & $95394.23$\\\hline
Perturbed Morse\footnote{ From regerence \cite{dk}}& $95394.23$\\\hline
nonlinear GHA & $89987.76$\\\hline

\end{tabular}
\caption{Values of the dissociation energy calculated for each model.}
\label{t0}
\end{table}

\begin{table}
\centering
\begin{tabular}{|c|c|c|c|c|c|}\hline
transition &\multicolumn{5}{|c|}{ energy ($cm^{-1}$)}\\ \cline{2-6}
 line& exp. & Morse & nl-GHA & linear GHA & perturbed Morse\footnote{There is a systematic difference of the $0.03cm^{-1}$ between the data used in references \cite{mwaz,hf} and that provided by the HITRAN database.}\\ \hline
$1\rightarrow0$  & 2143.24  &  2143.30    & 2144.56   & 2152.46 &  2143.27\\ \hline       
$2\rightarrow1$  & 2116.76  &  2116.39    & 2117.10   & 2123.32 &  2116.79\\ \hline       
$3\rightarrow2$  & 2090.34  &  2089.48    & 2089.99   & 2094.57 &  2090.37\\ \hline       
$4\rightarrow3$  & 2064.00  &  2062.57    & 2063.19   & 2066.21 &  2064.02\\ \hline       
$5\rightarrow4$  & 2037.72  &  2035.66    & 2036.68   & 2038.23 &  2037.73\\ \hline       
$6\rightarrow5$  & 2011.51  &  2008.75    & 2010.44   & 2010.63 &  2011.51\\ \hline       
$7\rightarrow6$  & 1985.38  &  1981.84    & 1984.43   & 1983.41 &  1985.35\\ \hline       
$8\rightarrow7$  & 1959.32  &  1954.93    & 1958.62   & 1956.55 &  1959.26\\ \hline       
$9\rightarrow8$  & 1933.33  &  1928.01    & 1932.97   & 1930.06 &  1933.24\\ \hline       
$10\rightarrow9$ & 1907.43  &  1901.10    & 1907.43   & 1903.93 &  1907.29\\ \hline       
$11\rightarrow10$& 1881.61  &  1874.19    & 1881.99   & 1878.15 &  1881.40\\ \hline       
$12\rightarrow11$& 1855.85  &  1847.28    & 1856.58   & 1852.72 &  1855.59\\ \hline       
$13\rightarrow12$& 1830.19  &  1820.37    & 1831.18   & 1827.63 &  1829.84\\ \hline       
$14\rightarrow13$& 1804.61  &  1793.46    & 1805.74   & 1802.89 &  1804.16\\ \hline       
$15\rightarrow14$& 1779.11  &  1766.55    & 1780.23   & 1778.48 &  1778.55\\ \hline       
$16\rightarrow15$& 1753.69  &  1739.64    & 1754.61   & 1754.39 &  1753.00\\ \hline       
$17\rightarrow16$& 1728.36  &  1712.73    & 1728.85   & 1730.64 &  1727.53\\ \hline       
$18\rightarrow17$& 1703.12  &  1685.82    & 1702.90   & 1707.21 &  1702.13\\ \hline       
$19\rightarrow18$& 1677.96  &  1658.91    & 1676.75   & 1684.09 &  1676.79\\ \hline       
$20\rightarrow19$& 1652.88  &  1632.00    & 1650.37   & 1661.29 &  1651.53\\ \hline       

\end{tabular}
\caption{Vibrational spectrum of the $\mathrm{CO}$ molecule. Comparison of the experimental data with the Morse, $q$-oscillator, perturbed Morse and GHA model. Experimental data from HITRAN database.}\label{t2}
\end{table}

\end{document}